\documentclass[12pt]{article}

\RequirePackage[OT1]{fontenc}
\usepackage{amsthm,amsmath,amssymb,amsfonts,bm,enumerate,xcolor,graphicx,natbib,color,url}
\RequirePackage[colorlinks,citecolor=blue,urlcolor=blue]{hyperref}

\def\frac#1#2{{\textstyle{#1\over#2}}}

\DeclareSymbolFont{AMSb}{U}{msb}{m}{n}
\DeclareMathSymbol{\Natural}{\mathbin}{AMSb}{"4E}
\DeclareMathSymbol{\Integer}{\mathbin}{AMSb}{"5A}
\DeclareMathSymbol{\Real}{\mathbin}{AMSb}{"52}
\DeclareMathSymbol{\Rational}{\mathbin}{AMSb}{"51}
\DeclareMathSymbol{\Imaginary}{\mathbin}{AMSb}{"49}
\DeclareMathSymbol{\Complex}{\mathbin}{AMSb}{"43} 
\DeclareMathSymbol{\Disk}{\mathbin}{AMSb}{"44} 
\def\bi{\begin{itemize}}
\def\ei{\end{itemize}}
\def\bd{\begin{description}}
\def\ed{\end{description}}
\def\ben{\begin{enumerate}}
\def\een{\end{enumerate}}
\def\bv{\begin{verbatim}}
\def\ev{\end{verbatim}}

\def\calC{{\mathcal C}}

\def\calN{{\mathcal N}}

\def\calS{{\mathcal{S}}}

\def\calT{{\mathcal T}}

\def\hat#1{{\widehat{#1}}}

\def\pr{{\rm Pr}}

\def\E{{\rm E}}

\def\cov{{\rm cov}}

\def\2to{{\ {\buildrel 2\over \longrightarrow}\ }}

\def\I1ton{{$I_1,\ldots,I_n$}}
\def\X1ton{{$X_1,\ldots,X_n$}}
\def\Y1ton{{$Y_1,\ldots,Y_n$}}
\def\Z1ton{{$Z_1,\ldots,Z_n$}}
\def\R1ton{{$R_1,\ldots,R_n$}}
\def\e1ton{{$e_1,\ldots,e_n$}}
\def\t1ton{{$t_1,\ldots,t_n$}}
\def\x1ton{{$x_1,\ldots,x_n$}}
\def\y1ton{{$y_1,\ldots,y_n$}}
\def\z1ton{{$z_1,\ldots,z_n$}}

\def\calS{{\mathcal{S}}}

\def\tt#1{{\texttt{#1}}}

\newcommand{\blind}{1}

\allowdisplaybreaks

\begin{document}
\thispagestyle{empty}
\baselineskip=28pt
\vskip 5mm
\begin{center} {\Large{\bf INLA goes extreme: Bayesian tail regression for the estimation of high spatio-temporal quantiles}}
\end{center}

\baselineskip=12pt

\vskip 5mm

\if1\blind
{
\begin{center}
{\large Thomas Opitz$^1$, Rapha\"el Huser$^{2\star}$, Haakon Bakka$^2$ and H\aa vard Rue$^2$}

\vspace{5pt}

{\small $^\star$Email: raphael.huser@kaust.edu.sa; Phone: +966 12 8080682}
\end{center}

\footnotetext[1]{
\baselineskip=10pt  INRA, UR546 Biostatistics and Spatial Processes, 228, Route de l'A\'erodrome, CS 40509, 84914 Avignon, France.}
}
\footnotetext[2]{
\baselineskip=10pt Computer, Electrical and Mathematical Sciences and Engineering (CEMSE) Division, King Abdullah University of Science and Technology (KAUST), Thuwal 23955-6900, Saudi Arabia.}\fi

\baselineskip=17pt
\vskip 4mm
\centerline{\today}
\vskip 6mm


\begin{abstract}
This work has been motivated by the challenge of the 2017 conference on Extreme-Value Analysis (EVA2017), with the goal of predicting daily precipitation quantiles at the $99.8\%$ level for each month at observed and unobserved locations. To propose a general approach to this specific problem, we develop a Bayesian generalized additive modeling framework tailored to estimate complex trends in marginal extremes observed over space and time. Our approach is based on a set of regression equations linked to the exceedance probability above a high threshold and to the size of the excess, the latter being modeled using the asymptotic generalized Pareto (GP) distribution suggested by Extreme-Value Theory. Latent random effects are modeled additively and semi-parametrically using Gaussian process priors, which provides high flexibility and interpretability. Fast and accurate estimation of posterior distributions may be performed thanks to the Integrated Nested Laplace approximation (INLA),  efficiently implemented in the \tt{R-INLA} software, which we also use for determining a nonstationary threshold based on a model for the body of the distribution.  We show that the GP distribution meets the theoretical requirements of INLA, and we then develop a penalized complexity prior specification for the tail index, which is a crucial parameter for extrapolating tail event probabilities. This prior concentrates mass close to a light exponential tail while still allowing heavier tails by penalizing the distance to the exponential distribution at a constant rate. We illustrate this new methodological framework through the modeling of spatial and seasonal trends in daily precipitation data provided by the EVA2017 challenge, comprising observations at $40$ stations in the Netherlands during the period 1972--2016. Capitalizing on \tt{R-INLA}'s fast computation capacities and  powerful distributed computing resources, we conduct an extensive cross-validation study to select model parameters governing the smoothness of trends, which are critical for accurate prediction. Our results clearly outperform simple benchmarks and are comparable to the best-scoring approach among the other teams. 
\end{abstract}

\baselineskip=16pt

\par\vfill\noindent


\newpage
\baselineskip=26pt

\section{Introduction}\label{sec:intro}
At its origins, Extreme-Value Theory focused on describing the asymptotic behavior of sample maxima \citep[see, e.g., the monograph of][]{Gumbel.1958}, and equivalent theoretical statements were later established for threshold exceedances \citep{Balkema.deHaan.1974,Pickands.1975}. Statistical methodology for threshold exceedances based on a limiting point process representation or on the related generalized Pareto (GP) distribution was developed by \citet{Hosking.Wallis.1987} and \citet{Davison.Smith:1990} among others, and numerous applications of threshold-based approaches to environmental extremes observed over space and/or time have since been published; see, e.g., \citet{Huser.Davison:2014}, \citet{Thibaud.Opitz.2015} and \citet{Bacro.etal:2017} for recent applications to precipitation data, \citet{ChavezDemoulin.Davison.2005} for temperature data, and \citet{Northrop.Jonathan:2011} and \citet{Jonathan.etal:2014} for oceanographic data. In contrast with the block maximum approach, the use of threshold exceedances allows detailed modeling of trends, seasonality and extremal clustering characteristics due to short-term dependence, while also giving more flexibility for balancing bias and variance; however, choosing a good threshold remains challenging \citep{Frigessi.al.2002,Scarrott.MacDonald:2012}, which has lead some authors to implement Bayesian algorithms accounting for threshold uncertainty \citep{Tancredi.al.2006}.

Fully Bayesian modeling approaches for spatial and/or temporal extremes often rely on latent processes embedded into the GP parameters to capture trends and dependence \citep{Cooley.al.2007,Bopp.Shaby.2017}. In particular, \citet{Cooley.al.2007} use Gaussian processes to capture spatial dependence and covariate-driven trends in precipitation data, taking advantage of simulation-based Markov chain Monte-Carlo (MCMC) methods for the estimation of posterior distributions. Here, we adopt a similar model structure for capturing spatio-temporal trends in precipitation data and for providing extreme quantile predictions. However, our statistical inference approach relies on the Integrated Nested Laplace Approximation (INLA) of posterior distributions, which is much faster and sidesteps convergence issues in simulation-based techniques, while providing highly accurate results \citep{Rue.al.2009,Martins.etal:2013,Rue.al.2017}. Furthermore, our choice of prior distributions allows us to appropriately smooth predicted quantiles over space and time, while borrowing strength across nearby space-time points, which is especially important when predicting rare events. We also derive the penalized complexity (PC) prior distribution for the GP tail index, which provides a principled prior choice \citep{Simpson.etal:2017} for this crucial parameter by penalizing the distance to a baseline model possessing a light, exponentially-decaying, tail.

Our strategy for modeling space-time trends in marginal precipitation extremes can be decomposed into three stages, each consisting of a suitable univariate response distribution combined with a regression equation capturing systematic temporal and spatial effects. The latter are described semi-parametrically in terms of appropriate Gaussian process priors. In the first modeling stage, we fit a Gamma regression to the positive precipitation intensities, in order to fix a suitable high spatio-temporal threshold. Next, we estimate the overall excess probability above this threshold through Bernoulli regression. Finally, we fit a GP distribution to the observed threshold exceedances, assuming a constant tail index and a scale parameter varying in space and time. We then predict extreme spatio-temporal quantiles by combining the posterior mean predictions from the Bernoulli and GP stages. For reasons of modeling and computational complexity, our Bayesian regression models are based on the working assumption that the data are conditionally independent with respect to the latent spatio-temporal trend components. By imposing this model simplification, the computational efficiency of INLA allows us to conduct an extensive cross-validation study for selecting certain crucial model parameters. 

This new methodology, whose development was motivated by the challenge organized for the 10th Extreme Value Analysis conference (EVA2017) held in June 2017 in Delft, Netherlands, has been applied to Dutch precipitation data, with the ultimate goal of predicting spatio-temporal quantiles at the $99.8\%$ level; see \citet{Wintenberger:2018} for more details on the dataset, the evaluation criterion, and the results of this competition.

In the remainder of the paper, Section~\ref{sec:data} presents the dataset and some preprocessing steps. The Bayesian space-time regression framework for tail modeling is developed in Section~\ref{sec:model}, where the derivation of the penalized complexity prior for the tail index is presented in Section~\ref{sec:pc}. We explain the Bayesian estimation approach using INLA in Section~\ref{sec:Inference}. The cross-validation study and our final results for the Dutch precipitation application are reported in Section~\ref{sec:applic}. Some concluding remarks with an outlook towards possible future developments are summarized in Section~\ref{sec:conclusion}.

\section{Data}\label{sec:data}
\begin{figure}[t!]
\centering
\includegraphics[width=\linewidth]{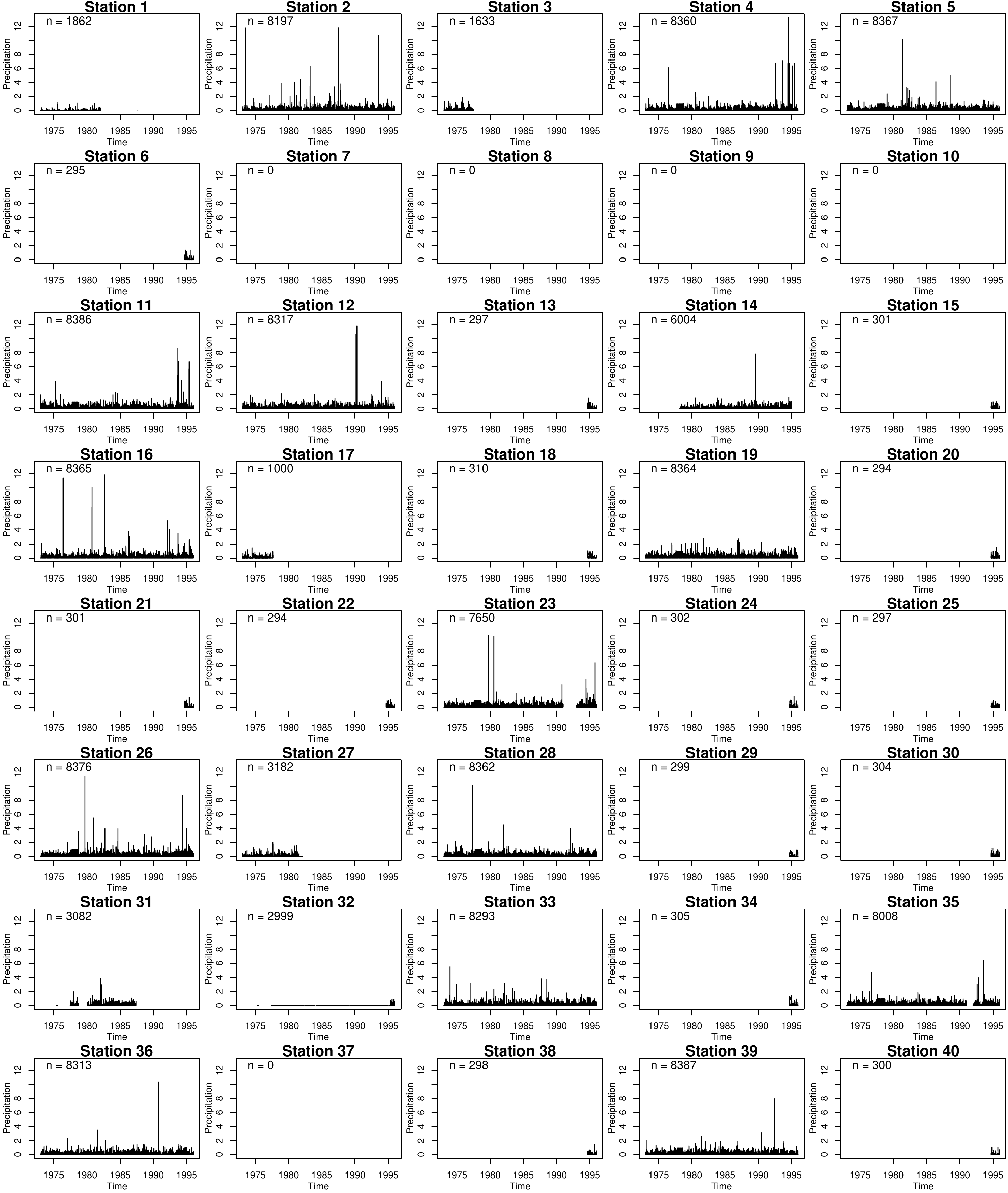}
\caption{Original precipitation time series measured at 40 stations during the training period. All time series are displayed using a common x-axis (time) and y-axis (precipitation amounts in inches). For each station, the sample size, ranging from $n=0$ to $n=8387$ for Station 39, is indicated in the top left corner of the corresponding panel.}\label{fig:data}
\end{figure}

The complete dataset consists of daily precipitation accumulations (in inches) measured at $40$ stations during the period 1972--2016. The data were divided into a training set (1972--1995) made available to the teams participating to the EVA2017 challenge, and a validation set (1996--2016) used to assess the high quantile predictions of the different teams, and therefore revealed only afterwards. Some of the stations were active during most of the training period, but many were not, resulting in a mixed dataset comprising a few time-rich stations and many time-poor ones; the available sample size varies from $n=0$ at Stations 6--10 and 37 to $n=8387$ at Station 39; see Figure~\ref{fig:data} and \citet{Wintenberger:2018} for more details on the data. In terms of modeling, this implies that it is crucial to build a spatial model that borrows strength across nearby stations to obtain robust and reliable predictions at unobserved, or scarcely observed, locations.

\begin{figure}[t!]
\centering
\includegraphics[width=0.8\linewidth]{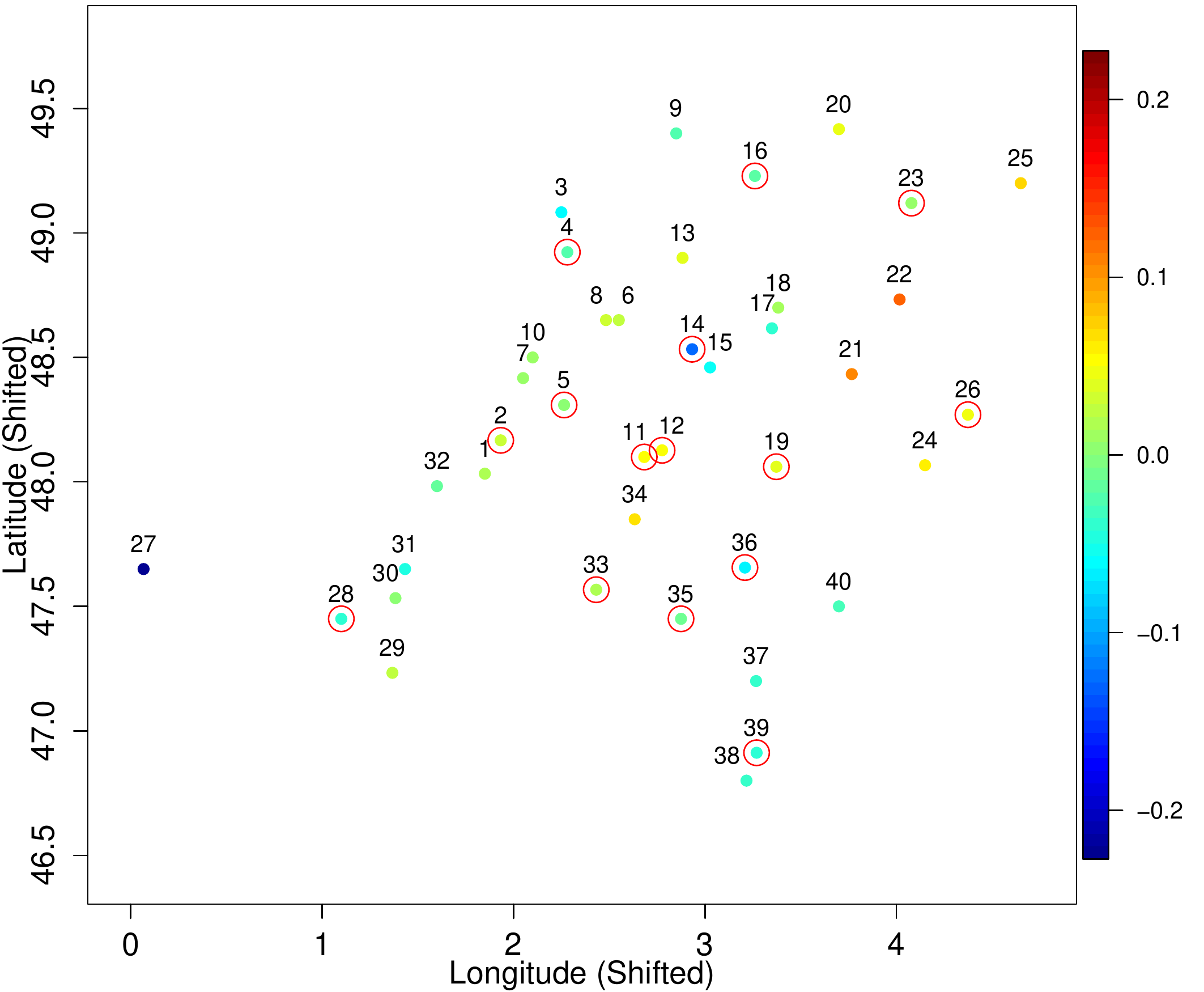}
\caption{Map of monitoring stations, plotted using the shifted coordinates provided by the EVA2017 challenge, and colored according to the estimated spatial random effect for Stage 1 (Gamma distribution) in our ``best model''; see \S\ref{sec:tailregression} and \ref{sec:results} for further details. Red circles indicate time-rich stations with at least $3650$ observations (i.e., about $10$ years of data) over the training period. \citet{Wintenberger:2018} displays the true locations and geographical map.}\label{fig:2}
\end{figure}

The exact coordinates of the stations were not disclosed before the end of the competition to prevent reverse engineering. However, shifted coordinates were provided and are displayed in Figure~\ref{fig:2}. As shown in \citet{Wintenberger:2018}, the region of study, revealed only afterwards, covers the Netherlands almost entirely. Although this country is quite small and mostly flat, which suggests that a stationary (or mildly non-stationary) process over space might be reasonable in this specific example, we opted to build a flexible Bayesian model for extreme events able to capture potentially complex trends in space and time; in this way, our general modeling and estimation approach may be applied to a wide range of scenarios. As Figure~\ref{fig:2} shows, time-rich stations are scattered across the study region, which is essential for borrowing strength across locations and efficiently estimating spatial trends.

Further exploratory plots (not shown) reveal that the distribution of precipitation varies slightly over time, with the strongest precipitation usually occurring during the summer. To capture this, our model described in \S\ref{sec:model} features a seasonal effect defined over weeks.

Figure~\ref{fig:data} also reveals that some parts of the data still contain erroneous measurements. For example, Stations 1 and 2 are geographically very close to each other, but their time series are very dissimilar; the data recorded in Station 1 appear in fact to be extremely small compared to the rest of the stations. Furthermore, Station 32 has a very long series of zeros from 1977 to 1995. Finally, Stations 4, 5, 11, 19, 23, 26, 28, 35 and 39 have a series of constant measurements during the period 1977--1978. We therefore removed all these highly suspicious data before proceeding further with the model fitting.

\section{Modeling}\label{sec:model}

\subsection{Asymptotic tail models}
In classical Extreme-Value Theory, the Fisher--Tippett--Gnedenko theorem characterizes  the convergence of renormalized maxima taken over a sequence of increasingly large blocks of independent and identically distributed (i.i.d.) random variables; see \citet{Coles.2001}, Chapter 3. The class of possible limit distributions may be summarized in terms of a single parametric family known as the generalized extreme-value (GEV) distribution. 
Equivalently, the generalized Pareto (GP) distribution may be used to model high threshold exceedances; see \citet{Coles.2001}, Chapter 4, and \citealp{Davison.Smith:1990}. The block maximum and threshold exceedance approaches can be unified using point process theory (\citealp{Coles.2001}, Chapter 7, \citealp{Davison.Huser:2015}). 
One assumes in practice that the asymptotic distribution provides a reasonable model for sample maxima or threshold exceedances for a fixed and finite block size or threshold, respectively. 
Provided that the asymptotic GP distribution holds \emph{exactly} to describe the tail of a random variable $Y\sim F$ above a predefined high threshold $u$, the upper tail may be represented in terms of the exceedance probability $p_u=\mathrm{pr}(Y>u)=1-F(u)$ and the GP distribution of positive threshold exceedance $Y_u^+={Y-u\mid Y>u}$, i.e.,
\begin{equation}\label{eq:gpd}
\pr(Y-u>y\mid Y>u) = {1-F(u+y)\over 1-F(u)} = 1-\mathrm{GP}(y;\sigma,\xi) = \left(1+\xi y/\sigma\right)_+^{-1/\xi},\quad y>0,
\end{equation}
where $a_+=\max(0,a)$, $\xi\in\mathbb{R}$ is the tail index and $\sigma>0$ is a scale parameter. When $\xi=0$, the expression in \eqref{eq:gpd} is interpreted as the limit $\xi\to0$ and corresponds to the exponential survivor function $\exp(-y/\sigma)$, $y>0$. When $\xi>0$, the GP distribution has a power-law decay with infinite upper endpoint, and when $\xi<0$, it has finite and parameter-dependent upper endpoint $-\sigma/\xi>0$, which leads to certain complications with likelihood-based inference. As our main goal is to model precipitation extremes, which are usually heavy-tailed, we restrict ourselves in this paper to $\xi\geq0$ and thus exclude the case $\xi<0$ below.

To avoid confounding problems due to the correlation between estimated GP parameters, it may be convenient to reparametrize the GP distribution using the tail index $\xi$ and a specific $q$-quantile $\kappa_q$ for some fixed probability of interest $q\in(0,1)$, i.e.,   
\begin{equation}\label{eq:gpd2}
\mathrm{GP}(y;\kappa_q,\xi)=\begin{cases} 1-\left[1+\{(1-q)^{-\xi}-1\}{y/\kappa_q}\right]_+^{-1/\xi}, & \xi\neq0,\\
1-(1-q)^{y/\kappa_q}, & \xi=0,
\end{cases}\quad y>0.
\end{equation}
In our Bayesian spatio-temporal analysis of extreme precipitation described in \S\ref{sec:applic}, we use the parametrization \eqref{eq:gpd2} in terms of the median. 

In practice, the equality between the left- and right-hand sides of \eqref{eq:gpd} should only be interpreted as an approximation for large $u$, and a careful bias-variance assessment must be performed to fix a suitable threshold $u$ above which observations are deemed to be \emph{extreme}; too low a threshold may cause sub-asymptotic bias, while too high a threshold implies large estimation variance due to the small number of threshold exceedances; see \citet{Davison.Smith:1990}, \citet{Northrop.Jonathan:2011} and \citet{Scarrott.MacDonald:2012} for standard diagnostics. In our application, we select $u$ by cross-validation; see \S\ref{sec:applic}.

We now present our general three-stage modeling strategy for estimating high quantiles, relaxing the above description to account for spatio-temporal trends in marginal extremes.

\subsection{Bayesian tail regression framework}\label{sec:tailregression}
The modeling of univariate extremes using the GP distribution has been popularized by \citet{Davison.Smith:1990}, who advocate to capture systematic variation in extreme events by including fixed covariate effects into the GP scale $\sigma$ (and potentially, if strongly supported by the data, also the tail index $\xi$). Similarly, using the equivalent point process representation of high threshold exceedances, \citet{Coles.Tawn:1996} propose to include spatial covariates into the extreme-value location and scale parameters, while treating the tail index as constant. With a different objective in mind, \citet{Beirlant.al.1999} formulate an exponential regression model for the (positive) tail index, and suggest that it may be adapted to incorporate exogeneous covariate information. To flexibly handle non-stationarity at high levels, \citet{Davison.Ramesh:2000} advocate a local likelihood approach for time series extremes, while \citet{ChavezDemoulin.Davison.2005} propose a generalized additive modeling (GAM) framework, whereby smoothing splines are incorporated into GP parameters. In the Bayesian framework, \citet{Casson.Coles.1999} and \citet{Cooley.al.2007} build hierarchical models for high threshold exceedances, which include latent spatial random effects  fitted using expensive simulation-based MCMC methods. 

The spatio-temporal Bayesian hierarchical approach that we develop here is similar in spirit to the purely spatial model of \citet{Cooley.al.2007}; however, the fast and accurate inference method based on INLA presented in \S\ref{sec:Inference} makes it possible to consider very high-dimensional data and  complex space-time models, comprising fixed and random effects that may potentially be organized hierarchically within the regression equations. 

Let $Y(s,t)$ denote daily precipitation observed at location $s\in\calS$ and time $t\in\{1,\ldots,n\}$, where $\calS\subset\Real^d$ is our study region. Our modeling and estimation strategy relies on the following three stages:
\begin{enumerate}
\item Positive precipitation intensities $Y_0^+={Y(s,t)\mid Y(s,t)>0}$ are first modeled assuming a Gamma distribution with mean $\mu(s,t)>0$ varying in space and time and constant shape parameter $k>0$, i.e., 
\begin{equation}\label{eq:gamma}
Y_0^+\sim {\rm Gam}\{y;\mu(s,t),k\}:={k^k\over \mu(s,t)^k \Gamma(k)} y^{k-1} \exp\{-k y/\mu(s,t)\},\qquad y>0,
\end{equation}
where $\Gamma(\cdot)$ is the Gamma function. For the purpose of using INLA, our parametrization of the Gamma distribution is slightly changed with respect to the classical one. A high space-time threshold $u(s,t)$ is then chosen as the $p_+$-quantile of positive precipitation intensities, deduced after fitting model \eqref{eq:gamma} to the data, where $p_+\in(0,1)$ is a fixed probability. The parameter $p_+$ defines the level of the threshold at each space-time location, and therefore controls the accuracy of the asymptotic GP approximation and the number of threshold exceedances in the following stages. As $p_+$ involves a crucial bias-variance trade-off, we selected it by cross-validation in \S\ref{sec:applic}.
\item Using the space-time threshold $u(s,t)$ obtained in Stage 1, exceedance indicators are modeled as Bernoulli random variates, i.e., setting $Z_u(s,t)=\mathbb I\{Y(s,t)>u(s,t)\}$,
\begin{equation}\label{eq:ber}
Z_u(s,t)\sim {\rm Ber}\{z;p_u(s,t)\}:=p_u(s,t)^z\{1-p_u(s,t)\}^{1-z},\qquad z\in\{0,1\},
\end{equation}
where $p_u(s,t)\in(0,1)$ denotes the \emph{overall} probability of exceeding the threshold, i.e., taking into account both the positive and zero parts of the precipitation distribution.
\item Using the space-time threshold $u(s,t)$ obtained in Stage 1, positive threshold exceedances $Y^+_u(s,t)={Y(s,t)-u(s,t)\mid Y(s,t)>u(s,t)}$ are assumed to follow the GP distribution parametrized in terms of its $q$-quantile $\kappa_q(s,t)$ and tail index $\xi\geq0$; recall \eqref{eq:gpd} and \eqref{eq:gpd2}. As the tail index usually lacks information for being accurately estimated, we treat it as constant over space and time. Moreover, assuming that the non-stationary patterns in the bulk and the tail of the distribution might be quite similar to each other, we further specify $\kappa_q(s,t)=\mu(s,t)r_q(s,t)$, where $\mu(s,t)$ is the Gamma mean estimated in the first stage using a rich training dataset, and $r_q(s,t)$ is a spatio-temporal correction. This allows to borrow strength from the bulk for more accurate tail estimation. Since we expect $r_q(s,t)$ to be close to stationary in practice, estimation is simplified in a Bayesian framework, where informative priors can be used. 
\end{enumerate}
The Gamma distribution in Stage 1 has been used a lot in hydrology as a whole statistical model for (positive) precipitation intensities \citep{Wilks:2006}. However, its tail decays exponentially fast, and is therefore often found to be too light and to underestimate probabilities associated with extreme events \citep{Katz.etal:2002}; by contrast, the GP model used in Stage 3 is motivated by asymptotic theory, and can capture heavy tails. Therefore, Stage 1 is only used here to select a suitable threshold $u(s,t)$ in a potentially highly non-stationary context and to get a reasonable prior for the size and shape of non-stationaries in the tail, while Stages 2 and 3 provide a complete characterization of the tail in terms of the threshold exceedance probability $p_u(s,t)$, the GP $q$-quantile $\kappa_q(s,t)$ and the tail index $\xi$. Note that Stage 1 is ``context-dependent'', in the sense that the Gamma distribution might need to be replaced by another distribution if the data were different, while Stages 2 and 3 are ``context-independent'', as Extreme-Value Theory holds under general assumptions. The overall $\alpha$-quantile $y_\alpha(s,t)$ of the precipitation distribution, for $\alpha>1-p_u(s,t)$, is then
\begin{align}
y_\alpha(s,t)&=u(s,t)+{\rm GP}^{-1}\{1-(1-\alpha)/p_u(s,t);\kappa_q(s,t),\xi\}\nonumber\\
&=\begin{cases}
u(s,t) + \kappa_q(s,t)\left[\{(1-\alpha)/p_u(s,t)\}^{-\xi}-1\right]/\left\{(1-q)^{-\xi}-1\right\},&\xi\neq0,\\
u(s,t) + \kappa_q(s,t)\log\{(1-\alpha)/p_u(s,t)\}/\log(1-q),&\xi=0,
\end{cases},\label{eq:highquantiles}
\end{align}
where ${\rm GP}^{-1}$ denotes the GP survivor function.

In order to capture systematic space-time variations in the data's tail behavior, we use a regression formulation with suitable link functions  and  an additive structure in the predictor components.  To keep the model fairly simple and robust while allowing for high flexibility, we assume that each  predictor includes (additionally to the intercept) a spatial random effect and a temporal random effect chosen to be separable, i.e., 
\begin{align*}
\log\{\mu(s,t)\} &=  \beta_0^{\mbox{\tiny Gam}}+x^{\mbox{\tiny Gam}}(s) + x^{\mbox{\tiny Gam}}(t),\\
{\rm logit}\{p_u(s,t)\} &= \beta_0^{\mbox{\tiny Ber}}+x^{\mbox{\tiny Ber}}(s) + x^{\mbox{\tiny Ber}}(t),\\
\log\{\kappa_q(s,t)\} &= \log\{\mu(s,t)\}+\beta_0^{\mbox{\tiny GP}}+x^{\mbox{\tiny GP}}(s) + x^{\mbox{\tiny GP}}(t),
\end{align*}
where $\beta_0^{\mbox{\tiny Gam}},\beta_0^{\mbox{\tiny Ber}},\beta_0^{\mbox{\tiny GP}}$ denote fixed intercepts, $x^{\mbox{\tiny Gam}}(s),x^{\mbox{\tiny Ber}}(s),x^{\mbox{\tiny GP}}(s)$ are spatial random effects defined at each station, and $x^{\mbox{\tiny Gam}}(t),x^{\mbox{\tiny Ber}}(t),x^{\mbox{\tiny GP}}(t)$ are temporal random effects defined on a weekly basis and cyclic with a yearly period. 
 The superscripts refer to each model stage. In our model, we include the logarithm of the Gamma mean $\log\{\mu(s,t)\}$ (or, in practice, of its estimated posterior mean) as an additive offset into the GP $q$-quantile, such that  $\log\{r_q(s,t)\}=\beta_0^{\mbox{\tiny GP}}+x^{\mbox{\tiny GP}}(s) + x^{\mbox{\tiny GP}}(t)$ can be interpreted as a residual effect after having removed the scaling implied by the Gamma distribution for positive precipitation. If the residual space and time effects $x^{\mbox{\tiny GP}}(s)$ and $x^{\mbox{\tiny GP}}(t)$, respectively, are not significantly nonzero, we can conclude that such trends in threshold exceedances are already well explained by the Gamma model. Since we adopt a Bayesian framework with Gaussian process priors centered at zero, including this offset also prevents an overly strong influence of the prior specification on posterior distributions when only few exceedances are observed. Routine calculations show that the GP density with the above link function is log-concave with respect to the predictor $\log\{\kappa_{q}(s,t)\}$, which is beneficial for INLA and optimization routines in general.

Each spatial effect, denoted by $x(s)$ for simplicity, is assumed to be driven by a zero mean Gaussian process prior with precision $\tau_s>0$ and Mat\'ern correlation function \citep{Stein:1999,Lindgren.etal:2011,Lindgren.Rue:2015}, i.e., 
\begin{equation}\label{eq:Matern}
\cov\{x(s_1),x(s_2)\}=\tau_s^{-1}{2^{1-\nu}\over \Gamma(\nu)}\left(\sqrt{2\nu}h/\psi\right)^\nu K_\nu\left(\sqrt{2\nu}h/\psi\right),\quad h=\|s_1-s_2\|,
\end{equation}
where $K_\nu$ denotes the modified Bessel function of second kind of order $\nu>0$, $\psi>0$ is the spatial range parameter, and $h$ is the distance between two locations $s_1,s_2\in\calS$. The spatial effect allows us to borrow strength across locations, its main purpose being to make predictions at unobserved (or scarcely observed) stations. Its small-scale behavior related to sample path regularity is of relatively minor importance here, and so we fix $\nu=1$. The range parameter $\psi$, however, controls the amount of spatial smoothing, and we select it by cross-validation in \S\ref{sec:applic} to optimize spatial prediction. Unlike $\psi$, the precision $\tau_s$ is estimated, rather than post-selected by cross-validation.

Each temporal effect, denoted by $x(t)$ for simplicity, is assumed to be driven by a second-order Gaussian random walk prior $x^\star(\omega)$ defined over weeks with precision $\tau_t>0$ for its weekly innovations (\citealp{Lindgren.Rue:2008} and \citealp{Rue.Held:2005}, Chapter 3). Let $\omega_t=\{1,\ldots,\lfloor n/7\rfloor\}$ denote the week associated to time $t\in\{1,\ldots,n\}$; one has $x(t)=x^\star(\omega_t)$, where, simultaneously for all weeks $\omega$,
\begin{equation}\label{eq:temporaleffect}
x^\star(\omega+1)-2x^\star(\omega)+x^\star(\omega-1)\sim \calN(0,\tau_t^{-1}),
\end{equation}
and the Gaussian process $x^\star(\omega)$ is restricted to sum to zero for identifiability and to be cyclic over a period of one full year (i.e., 52 ``weeks" with approximate length of $7$ days). Because the precision $\tau_t$ controls the smoothness of the weekly effect, we select it by cross-validation in \S\ref{sec:applic} to optimize temporal prediction.

Although the inference approach based on INLA may be exploited with additional fixed effects and more complicated random effects, potentially including space-time interactions, we restricted ourselves to the above structure, which was found to be flexible enough and to provide robust and interpretable results in our application. For the shape of the Gamma distribution, we fix a moderately informative Gamma prior distribution with shape $2$ and mean $1$, and due to the large sample size we can expect a negligible influence of the prior distribution on posterior estimates. The intercepts have noninformative Gaussian priors centered at $0$. The prior choice for the tail index is discussed in \S\ref{sec:pc}. Moreover, the priors for the various hyperparameters and additive regression components are mutually independent.

\subsection{Penalized complexity prior for the tail index}\label{sec:pc}
The tail index plays a crucial role in high quantile estimation. Light exponential tails arise with $\xi=0$, while heavier power-law tails arise with $\xi>0$. When $\xi\geq1$, the mean is infinite, and when $\xi\geq1/2$, the variance is infinite. Because of this, too large values of $\xi$ are unrealistic for many data types and typically lead to uncertainty inflation for high quantiles. It is therefore important to choose a suitable prior distribution for $\xi$, giving priority to light and moderately heavy tails while properly downweighting unrealistically heavy tails.

An elegant approach to choosing priors in a principled way when little expert knowledge is available consists in using the penalized complexity (PC) priors introduced by \citet{Simpson.etal:2017}, which penalize the ``distance'' from a base model at constant rate, independently of its parametrization. These priors  take the geometry induced by the choice of model parametrization into account, therefore avoiding intricate interpretation problems that arise otherwise. Furthermore, these priors are designed to allow for shrinkage towards a simpler reference model, which helps to prevent over-fitting by penalizing complex models. They also provide an ``objective'' (i.e., ``automatic'') way of choosing the prior distribution family, while keeping some degree of subjectivity in selecting the penalization rate parameter.

\begin{figure}[t!]
	\centering
	\includegraphics[width=\linewidth]{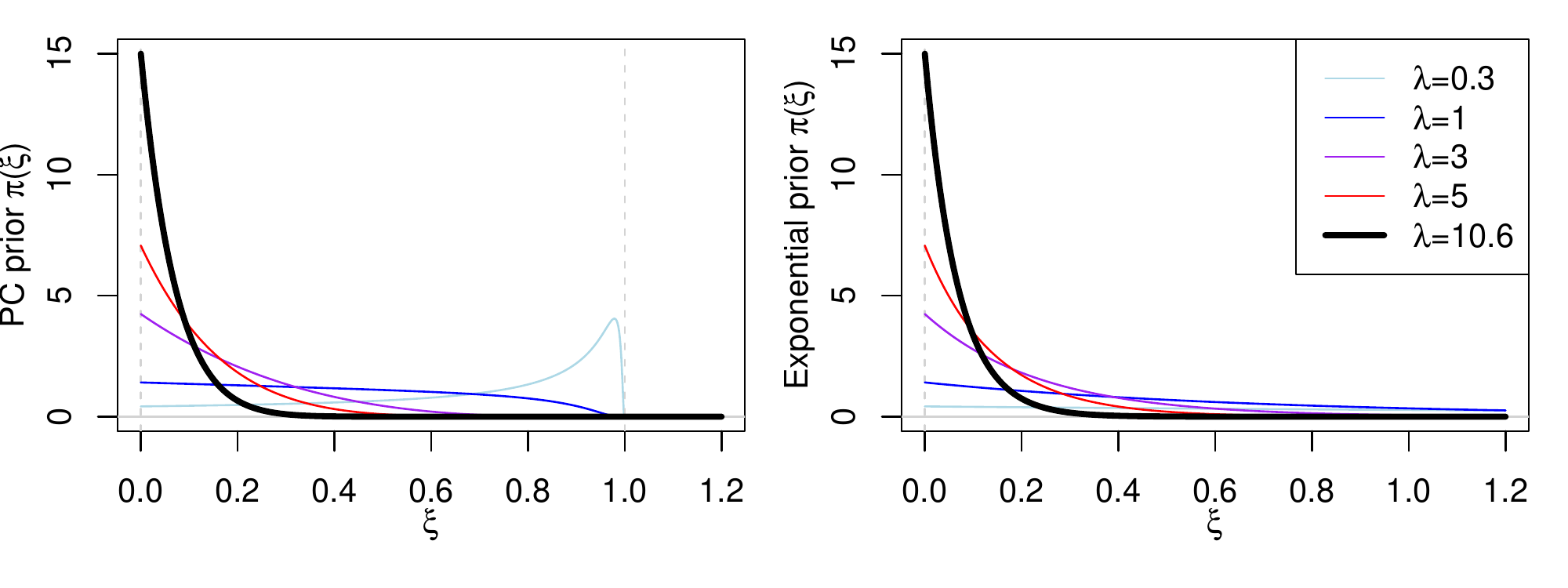}
	\caption{PC priors $\pi(\xi)$ for the GP tail index $\xi$ using the exact KLD formula in \eqref{PCpriorxi1} (left) or the approximate KLD formula in \eqref{PCpriorxi2} (right), i.e., using an exponential prior, setting the penalization rate parameter to $\lambda=0.3$ (light blue), $\lambda=1$ (dark blue), $\lambda=3$ (purple), $\lambda=5$ (red), and $\lambda=10.6$ (thick black), which was combined with \eqref{PCpriorxi2} in our application.}\label{fig:PCpriorsxi}
\end{figure}

Following \citet{Simpson.etal:2017}, a natural definition of the ``distance'' $d(f_\xi,f_{\xi_0})$ of a model $f_\xi$ with respect to a reference model $f_{\xi_0}$ (here assumed to be in the same family for simplicity) may be based on the Kullback-Leibler divergence (KLD) by setting 
\begin{equation}\label{KLD}
d(f_\xi,f_{\xi_0})=\sqrt{2\,\mathrm{KLD}(f_{\xi} || f_{\xi_0})},\quad\mbox{with}\quad \mathrm{KLD}(f_\xi || f_{\xi_0})=\int f_\xi(y)\log\left\{{f_\xi(y)/ f_{\xi_0}(y)}\right\}{\rm d}y.
\end{equation}
For the GP distribution \eqref{eq:gpd} with $0\leq\xi<1$, a natural choice of base model is the exponential distribution which arises when $\xi=0$. In this way, a prior assuming constant decay rate in terms of the Kullback-Leibler divergence \eqref{KLD} penalizes departure from an exponential tail, which usually makes sense for environmental data; the heavier the tail, the stronger the penalty. Let $f_\xi(y):=\sigma^{-1}\left(1+\xi{y/\sigma}\right)^{-1/\xi-1}$, $y>0$, $\sigma>0,\xi>0$, be the GP density and $f_{\xi_0}(y):=\lim_{\xi\to0}f_\xi(y)=\sigma^{-1}\exp(-y/\sigma)$, $y>0$, $\sigma>0,\xi_0=0$, be the exponential density. Knowing that the GP distribution \eqref{eq:gpd} with parameters $\sigma,\xi$ has mean $\sigma/(1-\xi)$ when $\xi<1$, and using the change of variable $t=\xi^{-1}\log(1+\xi y/\sigma)$, one obtains
\begin{align}
\mathrm{KLD}&(f_\xi || f_{\xi_0})=\int f_\xi(y)\log\left\{{f_\xi(y)/ f_{\xi_0}(y)}\right\}{\rm d}y\nonumber\\
&=\int_0^\infty \sigma^{-1}\left(1+\xi{y/\sigma}\right)^{-1/\xi-1}\log\left\{{\sigma^{-1}\left(1+\xi{y/\sigma}\right)^{-1/\xi-1}\over \sigma^{-1}\exp(-y/\sigma)}\right\}{\rm d}y\nonumber\\
&=\sigma^{-1}\int_0^\infty y\sigma^{-1}\left(1+\xi{y/\sigma}\right)^{-1/\xi-1}{\rm d}y-\left(1+{1\over\xi}\right)\int_0^\infty\log\left(1+\xi{y/\sigma}\right)\sigma^{-1}\left(1+\xi{y/\sigma}\right)^{-1/\xi-1}{\rm d}y\nonumber\\
&=\sigma^{-1}{\sigma\over1-\xi} - \left(1+{1\over\xi}\right)\xi\int_0^\infty t\exp(-t){\rm d}t={1\over1-\xi} - \left(1+\xi\right)\nonumber\\
&={\xi^2\over1-\xi},\qquad 0\leq \xi<1,\label{KLDGP}
\end{align}
which does not depend on the scale parameter $\sigma$, as expected. To define a PC prior $\pi(\xi)$ for the tail index $\xi$, we assume that a given model for the data $f_{\xi}$ is penalized at constant rate in terms of its ``distance'' $d(f_\xi,f_{\xi_0})=\sqrt{2\,\mathrm{KLD}(f_\xi || f_{\xi_0})}$ to the reference model $f_{\xi_0}$, therefore involving the exponential distribution in the ``metric'' space defined through $d(f_\xi,f_{\xi_0})$. Because the KLD \eqref{KLDGP} converges to infinity as $\xi\to1$, such a prior will put zero mass on $\xi\geq 1$, hence preventing infinite-mean models to occur. This can be seen as an advantage for applications in hydrology, where the tail index is usually quite small. We here propose two possible prior choices, which are based on (i) Equation \eqref{KLDGP} and (ii) an approximation of \eqref{KLDGP} as $\xi\to0$:
\begin{enumerate}
\item[(i)] The first possibility uses the \emph{exact} formula \eqref{KLDGP}, which yields $d(f_\xi,f_{\xi_0})=\sqrt{2}\xi/(1-\xi)^{1/2}$ and implies that the corresponding PC prior for $\xi$ with support $[0,1)$ is
\begin{eqnarray}
\pi(\xi)&=&\lambda\exp(-\lambda d(f_\xi,f_{\xi_0}))\left|{\partial d(f_\xi,f_{\xi_0})\over\partial \xi}\right|=\sqrt{2}\lambda\exp\left\{-\sqrt{2}\lambda{\xi\over(1-\xi)^{1/2}}\right\}\left\{{1-\xi/2\over(1-\xi)^{3/2}}\right\}\nonumber\\
&=&\tilde{\lambda}\exp\left\{-\tilde{\lambda}{\xi\over(1-\xi)^{1/2}}\right\}\left\{{1-\xi/2\over(1-\xi)^{3/2}}\right\},\quad 0\leq \xi<1,\label{PCpriorxi1}
\end{eqnarray}
where the penalization rate parameter is $\lambda=\tilde{\lambda}/\sqrt{2}>0$.
\item[(ii)] The second possibility is to approximate the KLD \eqref{KLDGP} by $\xi^2$, as $\xi\to0$. This yields $d(f_\xi,f_{\xi_0})=\sqrt{2}\xi$ and implies that the corresponding approximate PC prior for $\xi$, with support $[0,\infty]$, is exponential with rate $\tilde{\lambda}=\sqrt{2}\lambda>0$, i.e.,
\begin{equation}
\label{PCpriorxi2}
\pi(\xi)=\sqrt{2}\lambda\exp(-\sqrt{2}\lambda\xi)=\tilde{\lambda}\exp(-{\tilde\lambda}\xi),\quad \xi\geq0.
\end{equation}
\end{enumerate}
Our two proposed priors are illustrated in Figure~\ref{fig:PCpriorsxi}. 
These priors are very similar to each other when the penalization rate parameter $\lambda$ is large (giving strong preference to $\xi\approx0$) but they differ significantly when $\lambda$ is small. The PC prior using the exact KLD formula in \eqref{PCpriorxi1} gives zero probability to $\xi\geq 1$, while the prior \eqref{PCpriorxi2} gives exponentially decreasing but positive probability to all $\xi\geq 0$. This bounded/unbounded support distinction between the two proposed priors is more apparent for small $\lambda$; the PC prior \eqref{PCpriorxi1} becomes concentrated around $\xi=1$ as $\lambda\to0$, which is not very intuitive. In practice, the shrinkage rate $\lambda$ may be elicited from prior information in specific applications. For example, when the tail is known to be relatively light, $\lambda\geq 3$ seems to be a reasonable choice, and both priors should give similar results. In this case, both densities resemble an exponential density, which shrinks the tail index $\xi$ towards $\xi=0$; large values of $\xi$ will only arise if the data are heavy-tailed and provide strong evidence for it. In our precipitation example in \S\ref{sec:applic}, we use a fairly informative exponential prior for $\xi$ with rate $\tilde{\lambda}=15$ (i.e., $\lambda=15/\sqrt{2}\approx 10.6$), which sets the prior probability for tail indices larger than $\xi=0.2$ to be only about $5\%$; see Figure~\ref{fig:PCpriorsxi}.

\section{Bayesian inference}\label{sec:Inference}
\subsection{Posterior distributions}
The three regression models of our application are fitted separately, and the following exposition concentrates on  this approach.  It would be possible to have some of the fixed and random effects in common between the GP exceedance distribution and the Bernoulli distribution of exceedance indicators; in this case, we would fit those two regressions as a single model with two types of data/likelihood combinations. However, the Gamma regression must always be fitted separately in a first step, since it determines the data used for the other two regressions. By putting prior distributions on the fixed and random effects of the regression and on the various hyperparameters, we can apply Bayes' formula and infer the posterior distributions of model components of interest. 

In what follows, the notation is independent of the data/likelihood combination, given in each of the three stages as (1) positive precipitation intensities/Gamma, (2) exceedance indicators/Bernoulli and (3) threshold exceedances/GP, respectively.
For the observed data vector with components $y(s_i,t_i)$, $i=1,\ldots,m$, we write  $y=(y_1,\ldots,y_m)$ in short, and the vector of latent Gaussian predictors is denoted as $\eta=(\eta_i,\ldots,\eta_m)^T$  such that $\eta_i=\beta_0+x(s_i)+x(t_i)$. We denote hyperparameters related to the likelihood by the vector $\theta_y$, in our case given by the Gamma shape parameter and GP tail index, while the Bernoulli has no hyperparameter. The vector $x$ regroups all variables arising in the fixed and random effects, which includes $\beta_0$, weekly effects $x(t)$, which are based on a discretization of the year using $52$ variables,  and spatial effects $x(s)$ for $40$ observation and prediction sites $s$.  Hyperparameters related to the Gaussian priors of predictor components are collected in a vector $\theta_x$; they include the Mat\'ern precision $\tau_s$ and the precision $\tau_t$ of random walk innovations. The predictor vector $\eta$ is linearly related to the latent effects through an observation matrix $A$ determined by the structure of observation points and latent components, such that $\eta(x)=Ax$. The prior distribution on $x$ is Gaussian with zero mean and precision matrix (i.e., inverse covariance matrix) $Q(\theta_x)$. Owing to prior independence between the predictor components $\beta_0$, $x(s)$ and $x(t)$, $Q(\theta_x)$ is block-diagonal with three blocks. For parameters in $\theta_x$ and $\theta_y$ that are not fixed to a specific value, we define a prior distribution $\pi(\theta)$ with $\theta=(\theta_y,\theta_x)$, which is assumed to factorize over the components of $\theta$. The Gaussian density of $x$ is written $\pi(x\mid \theta_x)$. 

Using $\pi(y_i\mid \eta_i, \theta_y)$ to denote the likelihood of a data point $y_i$ conditional on its predictor $\eta_i$ and likelihood hyperparameters $\theta_y$, the joint density of $y$, $x$ and $\theta$ may be expressed as
\begin{equation}
\pi(y,x,\theta)=\pi(\theta)\times \pi(x\mid \theta_x)\times \prod_{i=1}^m\pi(y_i\mid \eta_i(x),\theta_y).
\end{equation}
For notational convenience, we now suppose that the predictor vector $\eta$ is a subvector of $x$ with $\eta_i=x_i$, $i=1,\ldots m$.
We are mainly interested in the posterior distributions of the variables in $x$ and $\theta$, whose joint posterior density conditional on the data is calculated through the Bayes formula as
\begin{equation}\label{eq:jointpost}
\pi(x,\theta\mid y) \propto \exp\left\{-{1\over 2}x^TQ(\theta_x)x+\sum_i \log \pi(y_i\mid x_i,\theta_y)+\log \pi(\theta)+\log K(\theta_x) \right\}
\end{equation}
where $K(\theta_x)$ is  a normalizing constant related to the hyperparameters of the latent Gaussian field. Specifically, we are interested in the univariate posterior distributions of predictors $\eta_i$ and components of $\theta$, which requires integrating the density \eqref{eq:jointpost} with respect to all components of $x$ and $\theta$ except the one of interest. Since we cannot provide an exact analytical solution of this high-dimensional integral, we could resort to simulation-based MCMC techniques \citep[see, e.g.,][Chapter 5]{Banerjee.al.2004} for producing a large representative sample of $x$ and $\theta$.  Instead, we propose using INLA, which exploits suitably chosen variants of analytical Laplace approximations of integrand functions \citep{Tierney.Kadane.1986}, such that integrals are over Gaussian densities and their evaluation becomes straightforward. 

\subsection{Integrated Nested Laplace Approximation (INLA)}
INLA \citep{Rue.al.2009,Rue.al.2017,Opitz.2017b} has become a cutting-edge tool for fast and accurate inference in complex and hierarchically structured GAM-like models with latent Gaussian structure and sparse precision matrix, such as our three models described above. For  a vector $v$ whose $k$th component has been removed, we write  $v_{-k}$. 
INLA  provides accurate approximations of the following two types of univariate posterior densities: 
\begin{align}
\pi(\theta_k \mid y) &= \int \pi(x, \theta \mid y)\,
\mathrm{d}x\,  \mathrm{d}\theta_{-k}, \label{eq:posthyper}\\
\pi(x_i\mid  y) &= \int\int \pi(x, \theta\mid y) \mathrm{d} x_{-i}\, \mathrm{d}\theta =  \int \pi(x_i \mid \theta, y)
\pi(\theta\mid y)\,\mathrm{d}\theta, \quad i=1,\ldots,\tilde{m} \label{eq:postpred}
\end{align}
where $\tilde{m}$ is the length of $x$. If a hyperparameter has been fixed to a specific value, then its prior density can be interpreted as a Dirac mass of $1$ at this value, and the related integral vanishes. In general, the number of estimated hyperparameters should be kept small, since integration with respect to components of $\theta$ is carried out numerically through an astute choice of discretization points. 
The main obstacle remains integration with respect to the Gaussian components $x$, for which the Laplace approximation is applied in a nested way: first, to approximate the posterior of the hyperparameters $\pi(\theta\mid y)$, and second to integrate with respect to the Gaussian vector $x_{-i}$. The log-concavity of the likelihood $\pi(y_i\mid \eta_i,\theta_y)$ with respect to $\eta_i$ is crucial since it ensures that a useful Gaussian approximation to $\pi(x\mid \theta,y)$ in \eqref{eq:jointpost} can be calculated by matching the mode and the curvature around it using an interative Newton--Raphson scheme based on  second-order Taylor expansions. Indeed, owing to the conditional independence of $y_i$ with respect to $\eta_i$, the precision matrix in the approximation is of the form $Q(\theta_x)+\mathrm{diag}(c)$, where the components $c_i$  in the added diagonal matrix must be positive and are closely related to the second derivative of $\log(y_i\mid x_i,\eta_y)$ with respect to $x_i$; hence the requirement of log-concavity. Notice that a simplified approximation scheme skips the second Laplace approximation by simply using a conditional Gaussian distribution for $x_{-i}\mid x_i,\theta$ based on the first Laplace approximation, but slight accuracy problems may arise with non-Gaussian univariate likelihoods.   For more details on the approximation mechanism and its implementation, see \citet{Rue.al.2009,Rue.al.2017} and \citet{Opitz.2017b}. 


\section{Results for the Dutch precipitation data application}\label{sec:applic}
\subsection{Cross-validation study}
The goal of the EVA2017 challenge was to predict the overall $\alpha_{\rm target}$-quantile of daily precipitation, with $\alpha_{\rm target}=0.998$, for each month and station $s\in \calC_j\subset\calS$, where $\calC_j$, $j=1,2$, denote two different sets of stations (so-called Challenges 1 and 2, respectively, see \citealp{Wintenberger:2018}). The target quantile is so high that fully non-parametric methods would likely perform quite badly and be outperformed by parametric alternatives; furthermore, we expect extreme-value models (such as the GP distribution \eqref{eq:gpd}) to provide a good fit at this level. The evaluation criterion used to rank the predictions submitted by all teams was based on the quantile loss function defined as
\begin{equation}\label{eq:quantileloss}
\ell_\alpha(y,q) = \begin{cases}
\alpha(y-q),& y>q,\\
-(1-\alpha)(y-q),& y\leq q,
\end{cases}
\end{equation}
where $y$ and $q$ represent observations and predicted quantiles, respectively. As the $\alpha$-quantile of the random variable $Y$ minimizes the expected risk function $q\mapsto\E\{\ell_\alpha(Y,q)\}$ \citep{Koenker:2005}, this justifies using \eqref{eq:quantileloss} for assessing quantile predictions, although alternative risk measures may also be used \citep{Daouia.etal:2018}. To optimize our spatio-temporal prediction performance at stations with poor time coverage, we select the most crucial hyperparameters, i.e., the Mat\'ern range parameter $\psi$ in \eqref{eq:Matern}, the temporal precision parameter $\tau_t$ in \eqref{eq:temporaleffect}, and the threshold probability $p_+$, using three different cross-validation (CV) criteria based on \eqref{eq:quantileloss}. An alternative would have been to estimate such parameters directly in the Bayesian model by specifying suitable prior distributions, but since we do not capture stochastic dependence in observations at nearby space-time points in our models, overfitting could have been an issue with such an approach.    As these hyperparameters determine the effective amount of information that is used or borrowed from neighboring space-time locations, they may strongly affect our predictions and thus need to be selected with care; recall that $p_+$ controls the bias-variance trade-off of the asymptotic GP approximation, $\psi$ controls the amount of spatial smoothing, and $\tau_t$ controls the amount of temporal smoothing of the corresponding random effects. Let $y_\alpha(s,t)$ denote the true $\alpha$-quantile for station $s$ and time $t$, and $\hat y_\alpha(s,t)$ denote our estimate of $y_\alpha(s,t)$ based on Equation~\eqref{eq:highquantiles}, obtained by fitting the three stage-model described in \S\ref{sec:model} to the entire dataset using INLA as explained in \S\ref{sec:Inference}. Let also $\hat y_\alpha^{-s}(s,t)$ and $\hat y_\alpha^{-j}(s,t)$ denote these predicted quantiles obtained after removing station $s$ and year $j$, respectively. We consider the following CV criteria designed to optimize spatial prediction, temporal prediction, or a combination of both:
\begin{align}
{\rm CV}_{\rm space}(\alpha)&=\sum_{s\in\calC_2}\sum_{j=1972}^{1995}\sum_{t\in\calT(j)}\ell_\alpha\{Y(s,t),\hat y_\alpha^{-s}(s,t)\},\label{eq:CV1}\\
{\rm CV}_{\rm time}(\alpha)&=\sum_{s\in\calC_2}\sum_{j=1972}^{1995}\sum_{t\in\calT(j)}\ell_\alpha\{Y(s,t),\hat y_\alpha^{-j}(s,t)\},\label{eq:CV2}\\
{\rm CV}_{\rm space-time}(\alpha)&={\rm CV}_{\rm space}(\alpha)+{\rm CV}_{\rm time}(\alpha);\label{eq:CV3}
\end{align}
here, $\calT(j)\in\{1,\ldots,n\}$ denotes the set of time points (days) within a specific year $j$. The lower the CV criteria the better the model. As the target probability $\alpha_{\rm target}=0.998$ is extreme, it may be better in practice to evaluate ${\rm CV}_{\rm space}(\alpha)$, ${\rm CV}_{\rm time}(\alpha)$ and ${\rm CV}_{\rm space-time}(\alpha)$ for some $\alpha<\alpha_{\rm target}$ in order to prevent overfitting. We tried to use $\alpha=0.995,0.998$ but without any major difference in the results. The results presented below and in \S\ref{sec:results} are based on $\alpha=\alpha_{\rm target}=0.998$. The final monthly quantile predictions were then extracted from our selected ``best model'' by averaging the daily predictions for that month.

After projecting the (shifted) longitude-latitude coordinates to a proper metric system, we considered $\psi\in\{25,50,75,100,125,150,175,200,225,250\}$km, $\tau_t=\sigma_t^{-2}$ with $\sigma_t\in\{0.005,0.010,0.015,0.020,0.025\}$, and $p_+\in\{0.90,0.91,0.92,0.93,0.94,0.95,0.96,0.97,0.98,0.99\}$, and fitted our three-stage hierarchical Bayesian tail regression model for all $500$ combinations of these hyperparameters. Notice that $p_+=0.99$ (for positive precipitation intensities) roughly corresponds to an overall probability of $\alpha=0.995$, as there were about $50\%$ of zeros observed at each station. Although model estimation based on INLA is much faster than classical simulation-based MCMC methods and it bypasses convergence assessment issues, the total number of fits to perform in order to compute the cross-validation criteria \eqref{eq:CV1}, \eqref{eq:CV2} and \eqref{eq:CV3} is equal to $29500$ ($500$ models times $59$ hold-out samples, corresponding to the sum of $35$ stations and $24$ years). In our case, a single fit (for all three stages) took about $8.5$ hours overall on $2$ cores, and we used the KAUST Supercomputer Shaheen II to compute them all in parallel. The total number of core-hours used was about half a million, which corresponds roughly to $57$ years of computation on a single-core machine. This cross-validation study underlines the importance of fast and accurate inference methods coupled with large distributed computing resources.

\begin{figure}[t!]
\centering
\includegraphics[width=\linewidth]{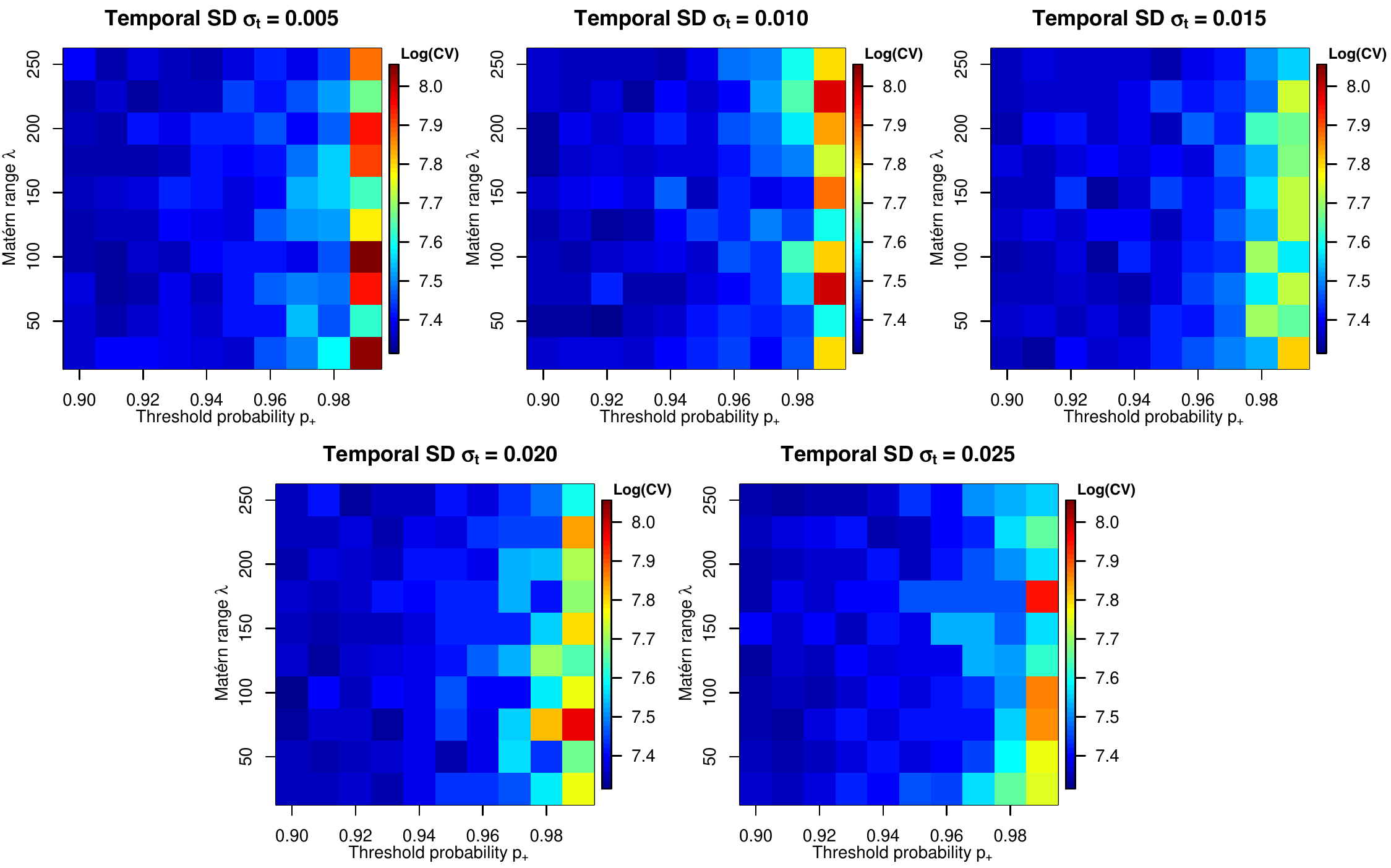}
\caption{Space-time cross-validation scores \eqref{eq:CV3} with $\alpha=0.998$, displayed on a common logarithmic color scale. Each panel displays the CV scores in terms of $\psi$ and $p_+$ for specific values of $\sigma_t$ (see panel titles). Lower values (i.e., darker blue cells) are better models.}\label{fig:CV}
\end{figure}

Figure~\ref{fig:CV} displays the space-time cross-validation scores \eqref{eq:CV3} for all models, while Table~\ref{tab:CV} presents a summary of the five best models. Overall, it appears that smaller probabilities $p_+$ yield much better predictions, thanks to the reduced variability due to the larger effective sample size of threshold exceedances. As for the other hyperparameters $\psi$ and $\sigma_t$, it is difficult to draw general conclusions, as the quite ``noisy'' diagnostics shown in Figure~\ref{fig:CV} make interpretation difficult. Nonetheless, the model that provides the best space-time predictions according to \eqref{eq:CV3} has $p_+=0.92$, $\psi=50$ and $\sigma_t=0.01$ (Model 122), and it yields reasonable results; hence, we decided to proceed with Model 122 as our ``best model''. As the models reported in Table~\ref{tab:CV} mostly yield a similar interpretation in terms of the estimated random effects, the next section showcases the results obtained for Model 122.

\begin{table}[t!]
\centering
\caption{Five ``best models'' ranked according to the space-time CV criterion \eqref{eq:CV3} with $\alpha=0.998$. Columns report the model ID, the Mat\'ern spatial range $\psi$, the temporal standard deviation (SD) $\sigma_t=\tau_t^{-1/2}$, the threshold probability $p_+$ for positive precipitation intensities, and the CV scores \eqref{eq:CV1}, \eqref{eq:CV2} and \eqref{eq:CV3}, respectively.}\label{tab:CV}
\vspace{5pt}
\begin{tabular}{c|ccc|ccc}
 & Range & Temp. SD & Proba. & \multicolumn{3}{c}{Cross-validation criteria}\\
Model ID & $\psi$ & $\sigma_t$ & $p_+$ & Space \eqref{eq:CV1} & Time \eqref{eq:CV2} & Space-time \eqref{eq:CV3}\\\hline 
$122$ & $50$ & $0.010$ & $0.92$ & $759.4$ & $751.9$ & $1511.3$\\
$304$ & $100$ & $0.020$ & $0.90$ & $760.9$ & $753.4$ & $1514.2$\\
$333$ & $75$ & $0.020$ & $0.93$ & $772.9$ & $747.3$ & $1520.2$\\
$211$ & $25$ & $0.015$ & $0.91$ & $767.3$ & $759.4$ & $1526.6$\\
$420$ & $250$ & $0.025$ & $0.91$ & $767.1$ & $759.6$ & $1526.8$
\end{tabular}
\end{table}

\subsection{Final results and interpretation}\label{sec:results}
Here, we provide further results and interpretation for our selected model. 
\begin{figure}[t!]
\centering
\includegraphics[width=\linewidth]{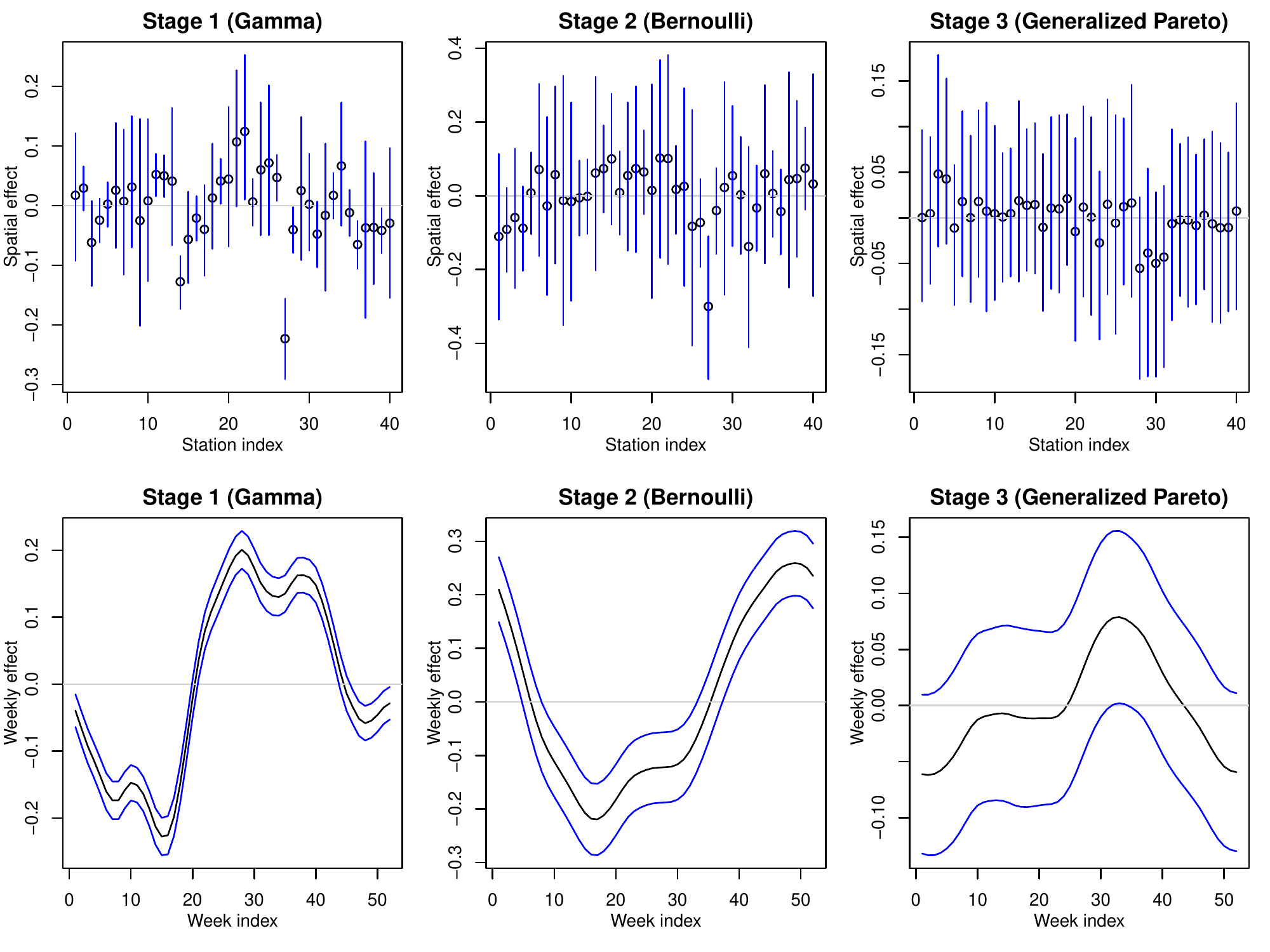}
\caption{Spatial (top) and weekly (bottom) effects displayed for the three stages (left to right) of our selected ``best model'' (Model 122 in Table~\ref{tab:CV}). Black dots and curves show the posterior means, while blue segments and curves are $95\%$ pointwise credible intervals.}\label{fig:Model122}
\end{figure}
Figure~\ref{fig:Model122} shows the posterior means and $95\%$ pointwise credible intervals estimated for the spatial and weekly random effects in each of the three stages (i.e., Gamma, Bernoulli and GP models), while Figure~\ref{fig:2} maps the spatial effect estimated in the first stage (Gamma). Overall, the uncertainty in the Gamma model is much lower than the Bernoulli and GP models, as it uses more information. Although these results neglect model selection uncertainty and model misspecification due to the conditional independence assumption of the data given the latent parameters, the weekly effect appears overall highly significantly different from zero for the Gamma model, while the GP model fitted to threshold exceedances (i.e., to the top $8\%$ of positive precipitation data) seems close to stationary in space and time when taking the uncertainty into account. As the GP scale parameter involves the Gamma mean as a multiplicative offset, this confirms that shape and size of non-stationary patterns are similar in the bulk and the tail of the distribution. The spatial patterns look quite stationary overall.

As expected, the credible intervals for the spatial effect in the Gamma model at time-rich stations are much narrower than those at time-poor stations; for example, Station 12, with $n=8317$ observations and surrounded by many other stations in the middle of the study region has the narrowest credible interval, while Station 9, with no observation and located on the Northern border of the study region has the widest credible interval. Interestingly, Stations 21 and 22, with very poor observational records, but surrounded by Stations 19, 23 and 26 with rich time series, have significant (or almost significant) positive effects, which confirms that our model succeeds in borrowing strength across nearby locations. Furthermore, the flexibility of our model is demonstrated by Station 27, located in the sea and far from the other stations, which stands out with a strong and highly significant negative effect; the corresponding time series shown in Figure~\ref{fig:data} indeed reveals that this particular station is less exposed to intense precipitation than stations inland. 

The weekly effect of the Gamma model is quite smooth and well estimated. It shows that precipitation intensity tends to be quite mild in April and stronger during summer, especially in July and September--October. The weekly effect of the Bernoulli model, however, has an opposite pattern and shows that the threshold $u(s,t)$ is more frequently exceeded from September to February than during the other months, which suggests that there are more wet days during the fall and winter, as expected. The GP weekly effect finally reveals that the largest exceedances over the threshold (and hence, the most intense precipitation events) tend to occur during the summer, though the associated uncertainty is high.
 
The GP tail index has posterior mean $\xi=0.34$, with $95\%$ credible interval $(0.31,0.38)$, which shows a significant departure from an exponential density (arising with $\xi=0$) and reveals that precipitation data in the Netherlands are heavy-tailed with finite first and second moments. This also suggests that the posterior distribution for $\xi$ is quite far from its prior displayed in Figure~\ref{fig:PCpriorsxi}; although the prior for $\xi$ was chosen to be quite informative with strong shrinkage towards light tails, the thousands of threshold exceedances available to fit the model provided strong evidence for heavier tails.

\begin{figure}[t!]
	\centering
	\includegraphics[width=\linewidth]{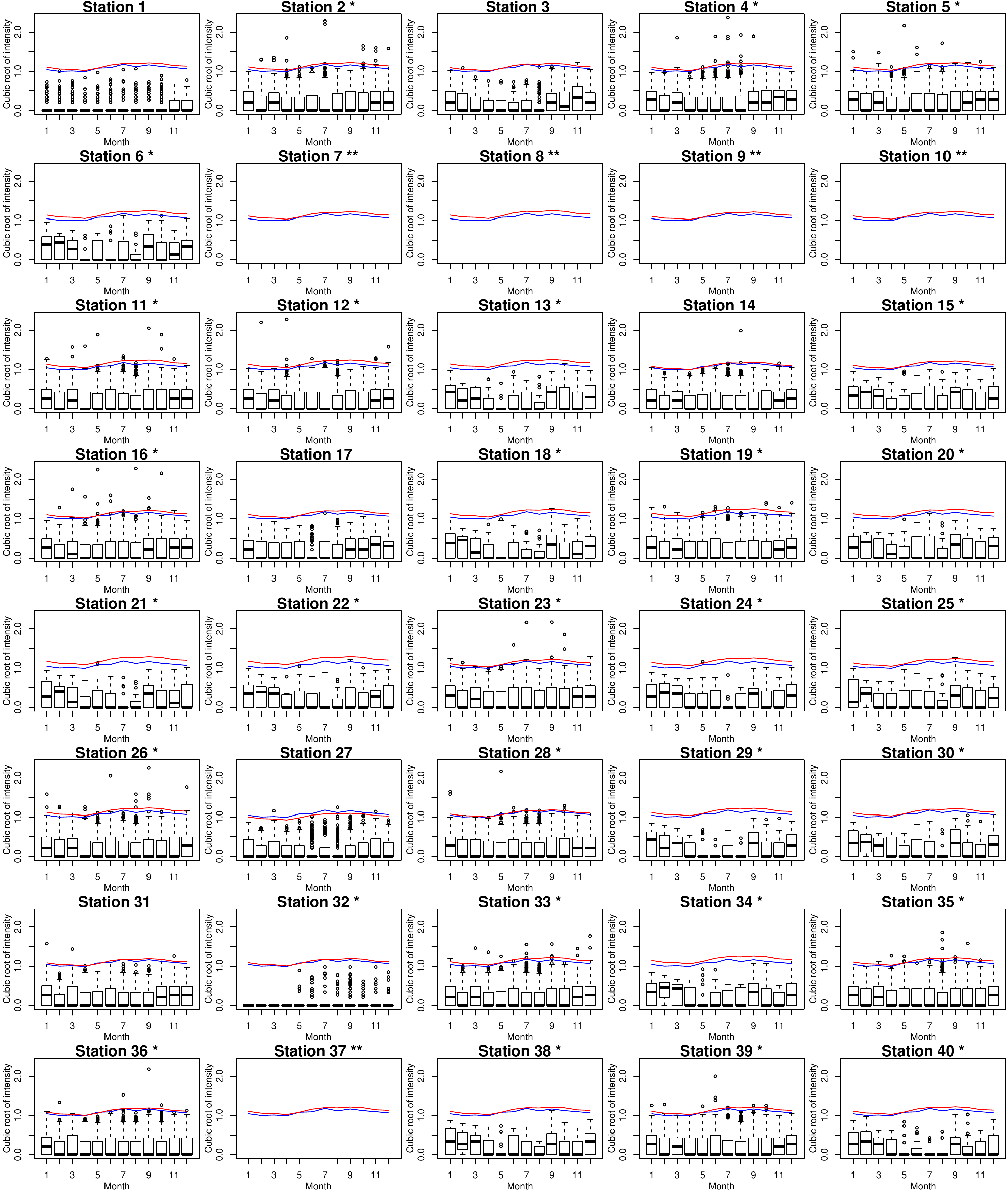}
	\caption{Boxplots of the cubic root of precipitation intensities for each month and station. Red curves are our model-based predictions of the $\alpha_{\rm target}$-quantile, with $\alpha_{\rm target}=0.998$, while blue curves are the empirical $\alpha_{\rm target}$-quantiles pooling all stations together. Stations, where prediction was required, are indicated by a $\star$ (Challenge 1) and $\star\star$ (Challenges 1 \& 2).}\label{fig:MonthlyFits}
\end{figure}
To conclude our analysis, Figure~\ref{fig:MonthlyFits} displays monthly boxplots of the data together with our final $\alpha_{\rm target}$-quantile predictions, for each month and station; the results are also compared to empirical quantiles computed for each month by pooling all stations together under the assumption of spatial stationarity. While our estimated weekly effect is quite weak but clearly visible, the difference between stations appears to be almost negligible, which suggests that we could have perhaps simplified the model by removing the spatial effect. Nevertheless, our Bayesian approach based on penalized complexity priors is designed to shrink the model towards a simple reference model, which helps in getting robust estimates and avoiding overfitting. Figure~\ref{fig:MonthlyFits} also illustrates the ability of our spatio-temporal modeling approach based on Extreme-Value Theory to extrapolate predictions to extreme quantiles, even at locations where no data have been recorded.

\section{Conclusion}\label{sec:conclusion}
Estimating extreme spatio-temporal quantiles from non-stationary data is not an easy task, and we think that the Extreme-Value Analysis conference 2017 challenge, which has motivated this work, has positively contributed to advancing the existing methods and the current literature. 

In this paper, we have combined the generalized Pareto distribution from Extreme-Value Theory with a flexible Bayesian latent Gaussian modeling approach. Our proposed three-stage model features spatial and temporal random effects that can capture systematic variations in the data. Our proposed generalized additive structures embedded in each stage provide interesting insight into the data's bulk and tail behaviors and could be made more complex if required by the context. Although the dataset studied for this competition is fairly small and weakly non-stationary, our very fast and accurate inference approach based on INLA could be applied to extremely high-dimensional and highly non-stationary space-time data with additional hierarchical levels.  Moreover, the combination of INLA and access to large distributed computing resources made it possible to run an extensive cross-validation analysis to select crucial hyperparameters and optimize our space-time quantile predictions. 

Despite the model complexity, our approach based on penalized complexity (PC) priors guarantees stable and robust quantile predictions and contributes to avoid overfitting, which is especially important when predicting extremes. In this paper, we have derived the PC prior for the tail index, which controls the tail decay rate and therefore highly impacts extrapolations to high levels.  Our proposed prior favors light exponential-like tail decay rates while penalizing unrealistically heavy tails.

Although the distribution in the first stage might need to be modified with other types of environmental data, our framework based on Extreme-Value Theory is very general and the methodology could be easily adapted to different contexts. Model estimation is executed using the \tt{R} package \tt{R-INLA}, which is convenient and easy to use. For reproducibility purposes, our code is illustrated on \url{http://www.r-inla.org/examples/case-studies}.

Although predicted quantiles from our model were quite good overall, certain aspects could have been improved. In particular, excluding months instead of years in our cross-validation study might have improved the estimation of the temporal effect, defined on a weekly basis. Furthermore, it would have been possible to cross-validate the tail index, as well, which is a crucial parameter for predicting extreme quantiles. However, because of our limited (though large) computational resources, we opted to perform our cross-validation study on parameters that control bias-variance trade-offs relevant for optimizing spatio-temporal prediction. Finally, we could have considered a more flexible model by allowing, for example, the tail index to vary over seasons or months, instead of being constant over space and time. However, this would increase the number of hyperparameters in the model, whose estimation might be more tricky and computer-demanding to perform using INLA.

In future work, it would be interesting to replace the generalized Pareto distribution by more flexible sub-asymptotic tail models \citep[see, e.g.,][]{Papastathopoulos.Tawn:2013,Naveau.etal:2016}, which could potentially be applied at much lower thresholds. Furthermore, as our model assumes that the data are conditionally independent given the latent parameters similarly to \citet{Cooley.al.2007}, it is not well-suited to properly capture space-time dependence, and it would be interesting to extend our methodological framework to more complex spatial \citep{Wadsworth.Tawn:2012,Thibaud.etal:2013,Thibaud.Opitz.2015,Huser.etal:2017,Castro.Huser:2017,Huser.Wadsworth:2018} and spatio-temporal \citep{Davis.etal:2013,Huser.Davison:2014,Bacro.etal:2017} models for extremes. However, such extremal models are much more difficult and expensive to fit, and when the primary goal is to estimate high \emph{marginal} quantiles as in this work, the exact characterization of the dependence structure is a secondary issue, or even perhaps a nuisance.


\section*{Acknowledgements}
The research reported in this publication was supported by funding from King Abdullah University of Science and Technology (KAUST). Support from the KAUST Supercomputing Laboratory and access to Shaheen II is gratefully acknowledged.

\baselineskip 20pt

\bibliographystyle{CUP}
\bibliography{biblio}

\baselineskip 10pt

\end{document}